\documentclass[twocolumn,epjc3]{svjour3}  

\journalname{Eur. Phys. J. A}

\usepackage{lipsum}
\usepackage{xcolor}
\usepackage{amsmath,amssymb} 
\usepackage{widetext}
\usepackage{hhline}
\usepackage{bbm}
\usepackage{diagbox}
\usepackage{rotating}
\usepackage{float}
\restylefloat{table}

\usepackage{cite}

\usepackage{graphicx}

\usepackage{subfigure}

\usepackage{tikz}
\usepackage{lipsum} 

\newcommand{\etal}{\textit{et al}. }
\usepackage[colorlinks=true, citecolor=blue]{hyperref}

\begin{document}
\title{ $A=4-7$ $\Xi$ hypernuclei based on interactions from  chiral effective field theory }
\author{Hoai Le\thanksref{addr1,e1}
\and Johann Haidenbauer\thanksref{addr1,e2}
\and Ulf-G. Mei{\ss}ner\thanksref{addr2,addr1,addr3,e3}
\and Andreas Nogga\thanksref{addr1,e4}
}
\thankstext{e1}{e-mail: h.le@fz-juelich.de}
\thankstext{e2}{e-mail: j.haidenbauer@fz-juelich.de}
\thankstext{e3}{e-mail: meissner@hiskp.uni-bonn.de}
\thankstext{e4}{e-mail: a.nogga@fz-juelich.de}

\institute{IAS-4, IKP-3 and JHCP, Forschungszentrum J\"ulich, D-52428 J\"ulich, Germany \label{addr1}
           \and
           HISKP and BCTP, Universit\"at Bonn, D-53115 Bonn, Germany \label{addr2}
           \and 
           Tbilisi State University, 0186 Tbilisi, Georgia \label{addr3}
}

\date{March 14, 2021}

\maketitle

\begin{abstract}
We investigate the  existence of bound $\Xi$ \break states in systems with $A=4-7$ 
baryons  using the Jacobi NCSM approach
in combination with  chiral NN and $\Xi$N interactions.  We find three shallow bound states for the NNN$\Xi$ system (with $(J^\pi,T)=(1^+,0)$,
$(0^+,1)$ and $(1^+,1)$) with quite similar binding energies. 
The $^5_{\Xi}\mathrm{H}(\frac{1}{2}^+,\frac{1}{2})$ and
$^7_{\Xi}\mathrm{H}(\frac{1}{2}^+,\frac{3}{2})$ hypernuclei are also clearly bound with respect to the 
thresholds $^4\mathrm{He} + \Xi$
and $^6\mathrm{He} +\Xi$, respectively. The binding 
 of all these $\Xi$ systems is predominantly due to the attraction 
 of the chiral $\Xi$N potential in the $^{33}S_1$ channel.
 A perturbative estimation suggests that the decay widths of all 
 the observed states could be rather small.

\end{abstract}

\section{Introduction}\label{sec:intro}
Recent progress in strangeness $S=-2$ nuclear physics \cite{Hiyama:2018lgs}, in particular 
the observation of nuclear bound states of 
$\Xi^{-} - ^{14}\mathrm{N}\, (^{15}_{\Xi}\mathrm{C})$ \cite{Nakazawa:2015joa,PhysRevLett.126.062501,Yoshimoto:2021ljs}
and possibly 
$\Xi^{-} - ^{11}\mathrm{B}\, (^{12}_{\Xi}\mathrm{Be})$
\cite{Nagae:2019uzt}, 
and evidence from femtoscopic measurements for an attractive $\Xi^{-}p$ interaction \cite{ALICE:2019hdt,ALICE:2020mfd}
have again triggered considerable interest in studying 
$\Xi$ hypernuclei theoretically 
\cite{Sun:2016tuf,Jin:2019sqc,Shyam:2019laf,Friedman:2021rhu,Kohno:2021str,Hu:2021ttg} 
despite  large uncertainties in the underlying $\Xi$N interaction.  
The latter is due to the absence of direct hyperon-hyperon (YY) and $\Xi$N scattering data, 
and the overall extremely scarce empirical information on $S=-2$ systems.  There are several $\Lambda\Lambda$ hypernuclei unambiguously determined in experiments, with the s-shell   $^{\text{ }\text{ }\text{ } \text{}6}_{\Lambda \Lambda}\text{He}$  \cite{PhysRevLett.87.212502,Nakazawa:2010zza}  being the lightest one. Given huge 
challenges in experimental identifications of 
 $\Xi$ states, the  possible existence  of s-shell $\Xi$ hypernuclei remains 
 by and large an open question.
Thus, theoretical predictions  for $\Xi$ hypernuclei especially for light systems are of great importance. Results from such studies can provide useful guidelines for experimentalists in searching for $\Xi$ bound states \cite{Ajimura:2019,Fujioka:2021S}.  

 In this work we explore the possible existence of light $\Xi$ hypernuclei 
 up to  $A=7$. For the lightest system, $^3_{\Xi}\mathrm{H}$, several calculations can be found in the literature \cite{Garcilazo:2016ylj,Filikhin:2017fog,Hiyama:2019kpw,Miyagawa:2021krh}.
 Not surprisingly, the predictions strongly depend on the interaction models used. 
 For example, Garcilazo \etal \cite{Garcilazo:2016ylj} and Hiyama \etal \cite{Hiyama:2019kpw} both obtained a deeply bound $^3_{\Xi}\mathrm{H}\,(J^\pi=3/2^+,T=1/2)$ state employing an effective $\Xi$N potential that mimics the phase shifts of the Nijmegen ESC08c potential \cite{Nagels:2015dia}. On the other hand, for interactions derived within chiral effective field theory (EFT)
 \cite{Haidenbauer:2018gvg} 
 or for $\Xi$N potentials deduced from lattice QCD 
 simulations by the HAL QCD Collaboration \cite{HALQCD:2019wsz}, 
 the system is found to be unbound \cite{Miyagawa:2021krh,Hiyama:2019kpw}.  
It has to be said that a strongly attractive $\Xi$N force as suggested by
that ESC08c potential, which even yields a $\Xi$N two-body bound
state \cite{Nagels:2015dia}, is not supported by the currently
available empirical constraints \cite{Haidenbauer:2015zqb}
including the aforementioned femtoscopic measurements, and also
not by lattice simulations at almost physical masses 
\cite{HALQCD:2019wsz}.
 Note that 
 in the work by Miyagawa \etal \cite{Miyagawa:2021krh} the original $\Xi$N interactions have been used directly to obtain the $t$ matrices entering the Faddeev equations for the three-baryon bound state. For the solution of these equations, however, only $\Xi$NN channels are considered.  
The variational calculation of Hiyama \etal \cite{Hiyama:2019kpw} 
is based on interactions where the various $YY$ channels of the 
original potential models \cite{Nagels:2015dia,HALQCD:2019wsz}
are renormalized into an effective $\Xi$N interaction and where 
 the latter is then treated within the so-called Gaussian
 expansion method \cite{Hiyama:2012sma}.
 Earlier studies of $\Xi$ hypernuclei with $A=5-12$ by Hiyama \etal 
 were performed within a cluster model \cite{Hiyama:2008fq}. 
 
  In the present study, we will employ the Jacobi no-core shell model (J-NCSM) in combination with microscopic nucleon-nucleon (NN) and 
  YY-$\Xi$N interactions derived within chiral EFT to investigate $A=4-7$ $\Xi$ hypernuclei.  
   Chiral EFT \cite{Epelbaum:2008ga} is a very powerful tool for precisely describing the NN interaction (see \cite{Reinert:2017usi} and references therein) and allows for accurate calculations of nuclear observables \cite{Epelbaum:2018ogq,Piarulli:2017dwd,Epelbaum:2019zqc,Maris:2020qne}. 
   It has also been successfully utilized in studies of the $\Lambda$N and $\Sigma$N 
   interactions by the
   J\"ulich-Bonn-Munich group \cite{Polinder:2006eq,Haidenbauer:2013oca,Haidenbauer:2019boi}. 
These chiral YN potentials have already been used to study $\Lambda$ hypernuclei within the J-NCSM approach
up to the $p$-shell \cite{Le:2019gjp,Le:2020zdu}.
Likewise, the YY-$\Xi$N potentials from chiral EFT up to  NLO~\cite{Haidenbauer:2015zqb,Haidenbauer:2018gvg} 
yield promising results for 
s-shell $\Lambda \Lambda$ hypernuclei \cite{Le:2021wwz}. 
  Therefore, it is very interesting to explore
  the predictions of chiral EFT for $\Xi$ hypernuclei, in 
  particular for light systems where a microscopic (\emph{ab initio}) treatment is possible. 
  
  The paper is organized as follows: in the next section we describe the baryon-baryon (BB) interactions employed in this work  focusing particularly  on the $S=-2$ BB potentials. Section~\ref{sec:JacobiNcsm} contains a brief description of the J-NCSM and its application to $\Xi$ hypernuclei, and 
  of the method to extrapolate the  binding and separation energies to infinite model spaces.
  In Section~\ref{sec:results}, our results for NNN$\Xi$, $^5_{\Xi}\mathrm{H}$ and $^7_{\Xi}\mathrm{H}$ are
  discussed. Final conclusions are given in Section~\ref{sec:Concl}. 
  
  %%%%%
  \section{Baryon-baryon interactions for $S=-2$}
  \label{sec:BBforces}
  
  For all calculations presented here, we employ BB interactions 
that are derived within chiral EFT \cite{Epelbaum:2008ga}. 
The high-order semilocal momentum-space regularized potential with a regulator of 
$\Lambda_{N}= 450$ MeV  (N\textsuperscript{4}LO{+}(450)) \cite{Reinert:2017usi} is used 
for the NN interaction.  
For the interaction in the $\Xi N$ channel, we employ the potential from
Ref.~\cite{Haidenbauer:2018gvg}. 
This interaction for the $S=-2$ sector has been constructed in agreement
with empirical constraints on the $\Lambda\Lambda$ $S$-wave 
scattering length and
with published values and upper bounds for $\Xi^- p$ elastic and
inelastic cross sections \cite{Haidenbauer:2015zqb}. Moreover, it yields
a moderately attractive $\Xi$-nuclear interaction as suggested by
experimental evidence for the existence of $\Xi$-hypernuclei
\cite{Nakazawa:2015joa,Yoshimoto:2021ljs}.
The value obtained for the $\Xi$ single-particle potential $U_\Xi(k=0)$ 
at nuclear matter saturation density is with around 
$-9$~MeV \cite{Kohno:2019oyw} noticeably smaller than the 
commonly cited potential depth of $-14$~MeV \cite{AGSE885:1999erv,Gal:2016boi}.
An application of this single-particle potential to finite 
$\Xi$ nuclei based on the local density approximation method
\cite{Kohno:2019oyw,Kohno:2021str}
showed, however, that pertinent predictions 
for the aforementioned recently reported states 
\cite{Nakazawa:2015joa,PhysRevLett.126.062501,Yoshimoto:2021ljs}
are quite in line with the energies observed in the
experiments.   

The considered $\Xi$N interaction includes the coupling to other 
BB channels in the strangeness $S=-2$ sector ($\Lambda\Lambda$, 
$\Lambda\Sigma$, $\Sigma\Sigma$).
However, in the actual calculation within the J-NCSM, it turned out that 
convergence of the eigenvalue iterations (that diagonalize the many-body Hamiltonian) 
to the lowest lying $\Xi$ states is rather poor when the coupling of
 $\Xi$N to $\Lambda\Lambda$ is explicitly included. Thus, similar to what
 has been done by Hiyama \etal \cite{Hiyama:2008fq,Hiyama:2019kpw}, the explicit 
 coupling to $\Lambda\Lambda$ is omitted. Instead, the contribution from
 the transition $\Lambda\Lambda -\Xi$N, which anyway can occur 
 only in the $^1S_0$ partial wave with isospin $I=0$,   
 is incorporated effectively by re-adjusting the strength of the
 corresponding $V_{\Xi \mathrm{N} -\Xi\mathrm{N}}$ potential.
 The other channel couplings in the $S=-2$ sector, i.e. 
 $\Xi \mathrm{N} - \Lambda\Sigma - \Sigma\Sigma$, are, however, still taken into account. In practice, the S-wave low-energy constants (LECs) 
 in the $I=0$ 
channel (to be concrete, those corresponding to the SU(3) singlet 
irrep. \{1\} \cite{Polinder:2007mp,Haidenbauer:2015zqb})
 are appropriately re-adjusted so that the real part of the $\Xi$N
 scattering length  remains practically the same after omitting the $\Lambda\Lambda$ channel.

Nonetheless, there is a delicate issue connected with that step. 
The chiral EFT interaction at NLO predicts the 
existence of a virtual state in the $I=0$, $^1S_0$ 
wave extremely close to the $\Xi$N threshold \cite{Haidenbauer:2015zqb,Haidenbauer:2018gvg}
-- a possible remnant of the 
$H$-dibaryon \cite{Haidenbauer:2011ah}. 
The virtual state is reflected in an impressive cusp structure in 
the $\Lambda\Lambda$ phase shift and a large $\Xi$N phase \cite{Haidenbauer:2015zqb}. 
The pertinent $\Xi$N scattering length is large too, 
cf.~Table~1 in Ref.~\cite{Haidenbauer:2018gvg},
with an imaginary part in the order of $10$~fm or more. 
Similar features are also seen in  
lattice QCD results \cite{HALQCD:2019wsz}. 
The large imaginary part somehow suggests an overall strong 
$\Lambda\Lambda-\Xi$N coupling. However, its value is artificially 
enhanced by the near-by virtual state. This can be easily seen 
by performing the
calculation in the particle basis and with physical masses. 
Then the cusp is strongly reduced \cite{Haidenbauer:2015zqb},
and the scattering lengths in the $\Xi^0$n channel 
amount to 
$a = (-1.30 -{\rm i}\,0.07)$~fm,
$(-2.05 -{\rm i}\,0.27)$~fm,
$(-1.95 -{\rm i}\,0.25)$~fm,
$(-1.41 -{\rm i}\,0.09)$~fm, respectively,
for the cutoffs $\Lambda=500$, $550$, $600$, and $650$~MeV 
in the regulator function
considered in \cite{Haidenbauer:2015zqb,Haidenbauer:2018gvg}.
Obviously, now the imaginary part is rather small and that means the
actual $\Lambda\Lambda-\Xi$N coupling is indeed fairly weak.
The latter conclusion has also been drawn in Ref.~\cite{Miyagawa:2021krh} where this issue has been  
examined from a slightly different perspective. 
See also the discussion in Ref.~\cite{Kohno:2021str}. 
     %------------------------------------------------------------------------------
          \begin{figure*}[htbp] 
      \begin{center}
      \hspace{0.3cm}{
     {\includegraphics[width=0.48\textwidth,trim={0.0cm 0.00cm 0.0cm 0 cm},clip]{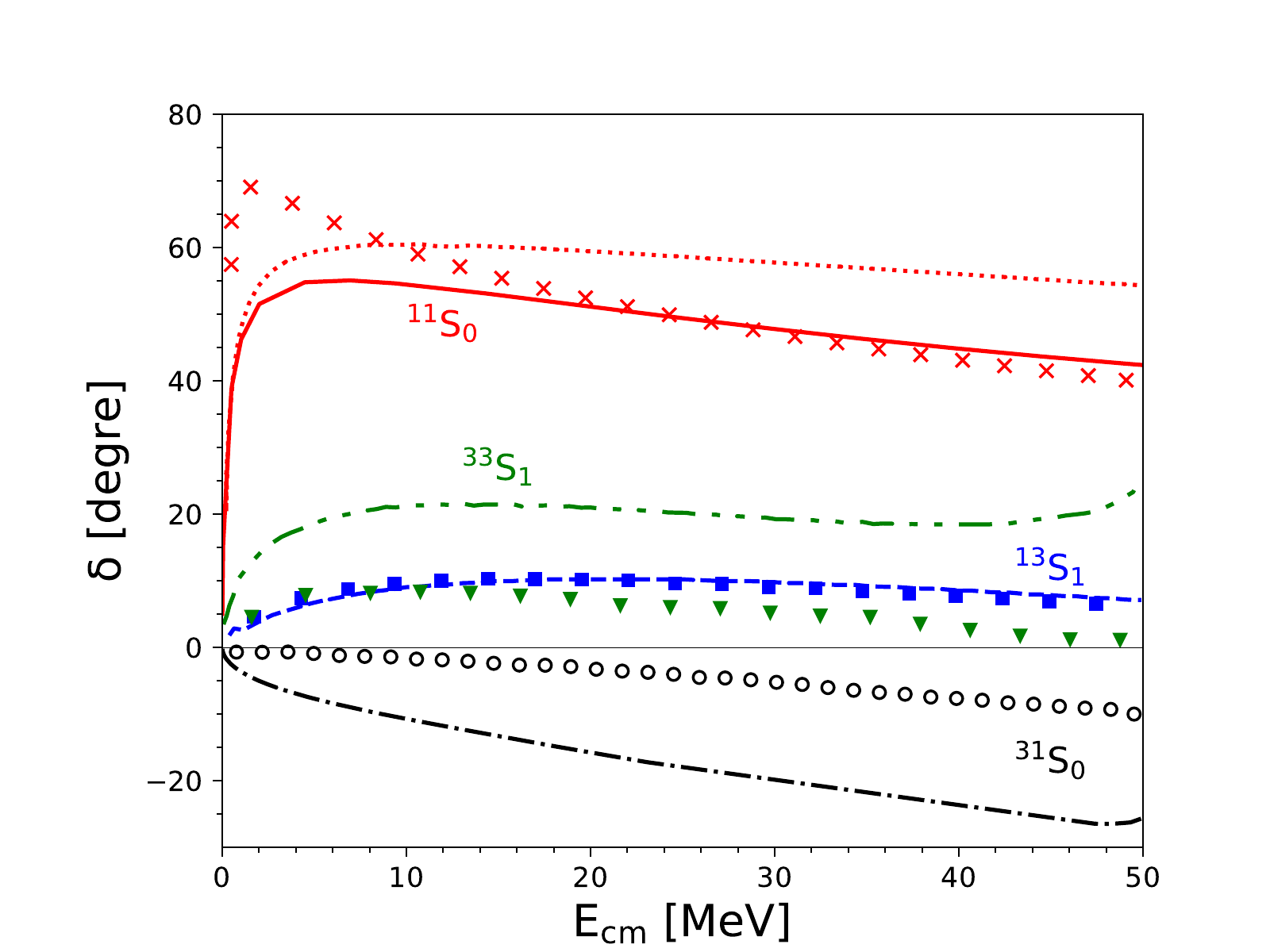}}
     { \includegraphics[width=0.48\textwidth,trim={0.0cm 0.00cm 0.0cm 0 cm},clip]{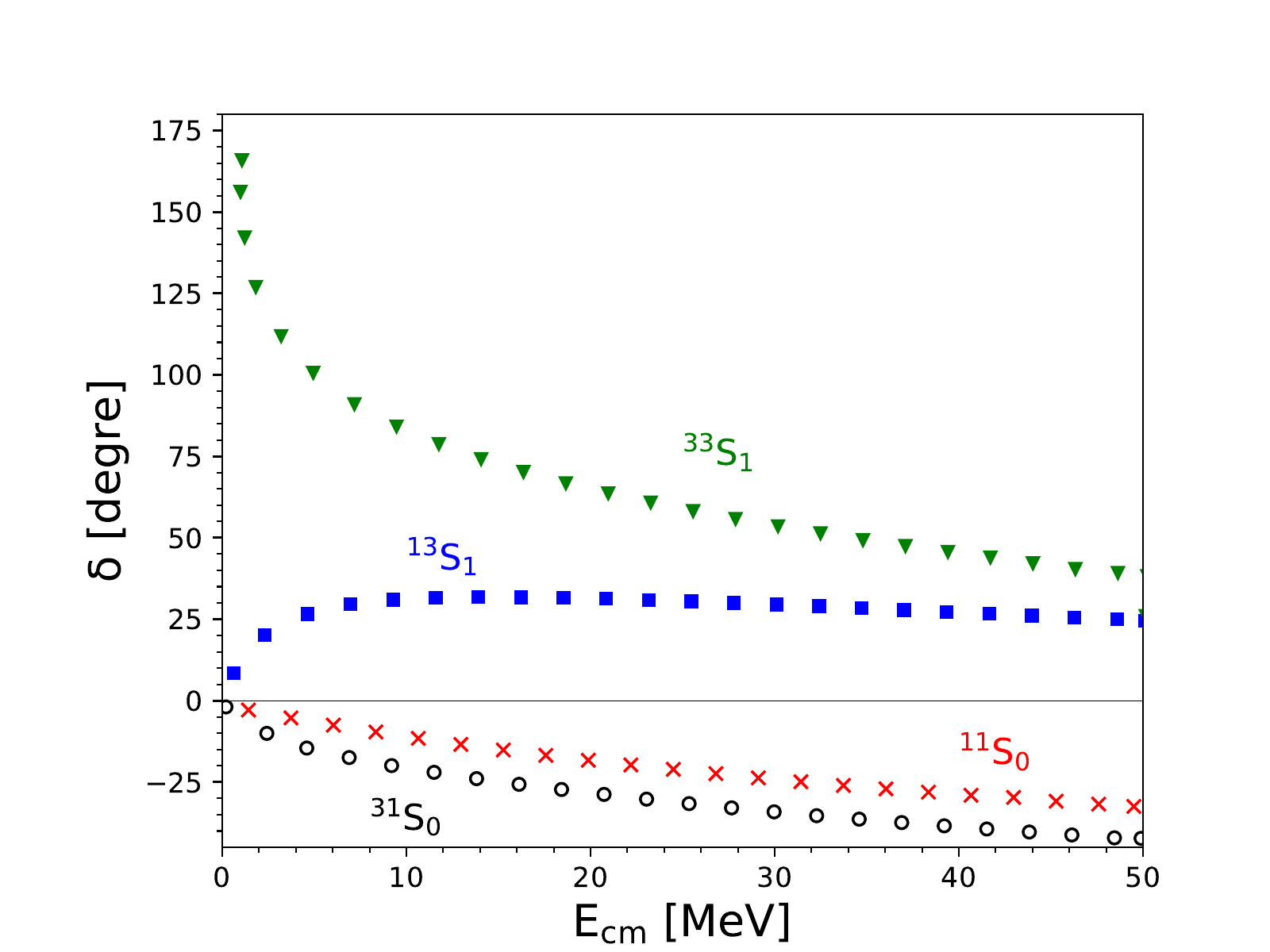}}}\\         
      \end{center}
                 \caption{$\Xi N$ phase shifts predicted by the 
                 NLO(500) and HAL QCD potentials (left panel) 
                 compared to those of the Nijmegen ESC08c model
                 (right panel). The NLO(500) results are shown
                 by lines: $^{11}S_0$ (dotted, red),  
                 $^{31}S_0$ (dash-dotted, black), $^{13}S_1$ (dashed, blue)
                 and $^{33}S_1$ (dash-double-dotted, green). The solid line
                 indicates the $^{11}S_0$ phase shift of the re-adjusted 
                 NLO(500) potential, see text. 
                 The HAL QCD and ESC08c results
                 (values are taken from \cite{Hiyama:2019kpw})
                 for $^{11}S_0$, $^{31}S_0$, $^{13}S_0$ and $^{33}S_0$ are indicated by crosses, circles, squares, and triangles,
                 respectively. 
                 Note the different scales in 
                 the left and right panels.}
    \label{fig:XiNphase_shift}
         \end{figure*}        
   %????
   
For the present exploratory study, we chose the YY-$\Xi$N NLO 
potential with 
the cutoff of $\Lambda =500$~MeV, i.e. the interaction with 
the smallest imaginary part. Moreover,
 we adjusted the LECs to the $\Xi^0$n scattering length, instead 
 of the one for $I=0$, in order
 to avoid any strong bias from the actual but unsettled location
 of the virtual state. After the re-adjustment the $^{11}S_0$ potential,  using the notation $^{(2I+1)(2S+1)}L_J$, 
 yields $a_{\Xi N}({^{11}S_0}) = -7. 00$~fm, to be 
 compared with $ a_{\Xi \mathrm{N}}({^{11}S_0}) = -7.71 -{\rm i}\,2.03$~fm 
 for the original potential where all coupled channels are 
 considered \cite{Haidenbauer:2018gvg}.
 We believe that this procedure allows us to capture the 
 essential features of the chiral $\Xi$N interaction in the 
 $^{11}S_0$ channel reliably, while guaranteeing at the same 
 time the applicability of the J-NCSM approach. 
 Note that all other $\Xi$N partial waves are not
 affected by this modification anyway
 and $\Lambda \Sigma$ and $\Sigma \Sigma$ components are included in the J-NCSM calculations. We, however, neglect YN interactions that are expected to give insignificant contributions but could potentially again induce $\Lambda \Lambda$ components to the many-body state. We postpone a more thorough investigation on this issue and 
 of the dependence of the $\Xi$ binding energies  
 on the chiral cutoff $\Lambda_{YY}$ to a future study. 

 Finally,  to speed up the convergence of the J-NCSM calculations, the NN and YY interactions are evolved using the similarity renormalization group (SRG) \cite{Bogner:2007hn}. Thereby, we use 
 an SRG flow parameter of  $\lambda_{NN} =1.6$~fm\textsuperscript{-1} for the NN interaction. This value has already been used in Refs.~\cite{Le:2020zdu,Le:2021wwz} and is motivated by the observation that ordinary nuclei are bound fairly realistically even if three-nucleon forces are neglected for this $\lambda_{NN}$. 
 The $S=-2$ potential 
 is SRG-evolved to a wide range of  SRG  flow parameter (denoted generically by $\lambda_{YY}$), namely   
 $1.4 \leq \lambda_{YY}  \leq 3.0$~fm\textsuperscript{-1}.
 The variations of the binding energies with respect to $\lambda_{YY}$ allow one to 
  quantify  the possible contribution of  the omitted SRG-induced three- and more-body forces. Note that such contributions are remarkably small for $\Lambda\Lambda$ hypernuclei \cite{Le:2021wwz}. 
   
 It is worthwhile to compare the $\Xi$N phase shifts of the employed 
 EFT interaction NLO(500) with those predicted by 
 the Nijmegen ESC08c \cite{Nagels:2015dia} and the HAL QCD \cite{HALQCD:2019wsz} potentials. As mentioned in the introduction, the 
 latter two interactions have recently been considered in $A=3,4$
 $\Xi$~hypernuclear calculations by Hiyama \etal \cite{Hiyama:2019kpw}.  
 The phase shifts for the four S-wave states, namely $^{11}S_0$, $^{31}S_0$,  $^{13}S_1$ and $^{33}S_1$,
 are displayed in Fig.~\ref{fig:XiNphase_shift}. As expected, the 
 original NLO(500) interaction (cf. the dotted line) and the 
 re-adjusted potential differ only slightly in 
 the $^{11}S_0$ phase shifts.
 Overall, the results by the NLO(500) and HAL QCD interactions 
 are fairly similar to each other, but differ substantially from the Nijmegen ESC08c potential. 
 The ESC08c is strongly attractive in the $^{33}S_1$ channel 
 (leading to a deuteron-like $\Xi$N bound state), whereas the 
 chiral NLO(500) (HAL QCD) interaction is only moderately (weakly)
 attractive in this channel.  
 Moreover, while the $^{11}S_0$ $\Xi $N interaction is rather attractive 
 in the HAL QCD and NLO(500) potentials, it is actually repulsive 
 in the ESC08c model.  
 Although the NLO(500) and HAL QCD $\Xi $N phase shifts exhibit an overall 
 rather similar trend, there are visible differences in all $\Xi N$
 partial waves except for $^{13}S_1$. As we will 
 discuss later, such variations lead to qualitative differences
 in the predictions of the two interactions for light $\Xi$ systems.
 
  \section{Jacobi NCSM for $\Xi$ hypernuclei}\label{sec:JacobiNcsm}
  The application of  the Jacobi no-core shell model (J-NCSM) to $\Xi$ hypernuclei follows very closely our J-NCSM formalism for 
  $\Lambda\Lambda$ systems described in   \cite{Le:2021wwz}.   Here we also split the basis functions into two orthogonal sets:
  one set that involves two $S=-1$ hyperons, \\ $ |\alpha^{*(Y_1 Y_2)}\rangle $, and the other that contains the doubly
  strange $\Xi$ hyperon,  $|\alpha^{* (\Xi)}\rangle$.   The  $|\alpha^{* (\Xi)}\rangle $ states are exactly the same as  constructed in \cite{Le:2021wwz},
      \begin{eqnarray}\label{eq:basisS=Xi}
&             |\alpha^{* (\Xi)}\rangle & =  |\alpha_{(A-1)N}\rangle \otimes  |\Xi\rangle\nonumber\\[2pt]
&   & = | \mathcal{N}{J}{T}, \alpha_{(A-1)N}\, n_{\Xi}\, I_{\Xi} \,t_{\Xi} ; \nonumber\\
& & \quad  (J_{A-1} (l_{\Xi}\, s_{\Xi})\,I_{\Xi}){J}, 
(T_{A-1}\, t_{\Xi}) {T} \rangle\,.
\end{eqnarray}
  Since the $\Lambda\Lambda -\Xi $N transition is absorbed into the strength of the $V_{\Xi \mathrm{N} -\Xi N}$ potential and since we omit YN interactions, the basis states   $ |\alpha^{*(Y_1 Y_2)}\rangle $
 can be  restricted to  
  \begin{eqnarray}\label{eq:basisS=Y1Y2}
 &   |\alpha^{*(Y_1 Y_2)}\rangle    &    =  |\alpha_{(A-2)N} \rangle  \otimes | Y_1  Y_2\rangle\nonumber\\[2pt]
&      & = | \mathcal{N}{J}{T}, \alpha_{(A-2)N}\, \alpha_{Y_1 Y_2}\, n_{\lambda} \lambda;\nonumber \\[2pt]
 & & \qquad   ((l_{Y_1 Y_2} (s_{Y_1} s_{Y_2}) S_{Y_1 Y_2}) J_{Y_1 Y_2} 
(\lambda J_{A-2})I_{\lambda}) J,\nonumber\\[2pt]
&  & \qquad  ((t_{Y_1} t_{Y_2})T_{Y_1 Y_2} T_{A-2}) T \rangle ,
\end{eqnarray}
 with $|Y_1 Y_2 \rangle =| \Lambda \Sigma\rangle $  or  $|\Sigma \Sigma\rangle$. The notations  used in Eqs.~(\ref{eq:basisS=Xi},\ref{eq:basisS=Y1Y2})
 are the same as in  \cite{Le:2021wwz}\footnote{To be consistent with 
 \cite{Le:2021wwz} we use here and later on $t,\,T$ for the isospins}.  
 For example, the symbol $\alpha_{(A-2)\mathrm{N}}$
  stands for all  quantum numbers characterizing the  antisymmetrized  states of $A-2$ nucleons:
the total number of  oscillator  quanta $\mathcal{N}_{A-2}$, total angular momentum $J_{A-2}$, isospin $T_{A-2}$ 
and state index $\zeta_{A-2} $ as well. Similarly,  $\alpha_{Y_1 Y_2}$ stands for a complete set of  quantum numbers describing the subcluster of two hyperons $Y_1$ and $Y_2$.  The principal quantum  number $n_{\lambda}$ of the
harmonic oscillator (HO) together with the orbital 
 angular $\lambda$  describe the  relative motion of the $(A-2)$N core with respect
to the center-of-mass (C.M.)~of the $Y_1 Y_2$ subcluster.   The orders, in which these quantum numbers are coupled, are  shown after the semicolon.
For practical calculations, we truncate the model space by limiting the total HO energy quantum number $\mathcal{N}$
in Eqs.~(\ref{eq:basisS=Xi},\ref{eq:basisS=Y1Y2}) to $\mathcal{N} \leq \mathcal{N}_{max}$ where $\mathcal{N}= \mathcal{N}_{A-2} + 2n_{\lambda} + l_{\lambda} + \mathcal{N}_{Y_1 Y_2 }$ or  $\mathcal{N}=\mathcal{N}_{A-1}  + l_{\Xi} + 2n_{\Xi}  $, respectively. As a consequence, the computed binding energies
will depend on $\mathcal{N}_{max}$ and on the HO frequency $\omega$. To extract the converged results we will follow the two-step
extrapolation procedure that has been successfully  employed for nuclear and hypernuclear energy calculations  \cite{Liebig:2015kwa,Le:2020zdu,Le:2021wwz}. First,  the energies $E(\mathcal{N}, \omega)$ are computed for all accessible model spaces $\mathcal{N}_{max}$ and for a wide range of  $\omega$. Then 
$E_{\mathcal{N}}$ is determined for a given $\mathcal{N}_{max} = \mathcal{N}$ 
by minimizing the energies $E(\mathcal{N}, \omega)$ with respect to $\omega$. In the second step, an exponential fit is applied to $E_{\mathcal{N}}$ in order to extrapolate to $\mathcal{N} \to \infty$. 

Furthermore,   in order to  write down the explicit form of the Hamiltonian, we  also distinguish three parts of the Hamiltonian,   namely  
$H_{Y_1 Y_2},  H_{\Xi}$ and $H^{S=-2}_{Y_1Y_2, \Xi \mathrm{N}}$, like for $\Lambda\Lambda$ hypernuclei. As mentioned before, we do not 
take into account  YN interactions in the $S=-1$ sector here. 
 Hence, the $H_{Y_1 Y_2}$,   $H^{S=-2}_{Y_1Y_2, \Xi \mathrm{N}}$  and $ H_{\Xi}$  can be written as
\begin{eqnarray}  \label{eq:hamiltonian2Y}
&  & H_{Y_1 Y_2}   = \sum_{i < j=1}^{A-2} \Big( \frac{2p^{2}_{ij}}{M(t_{Y_1}, t_{Y_2})} \,+ V^{S=0}_{ij} \Big)\nonumber\\[2pt]
 & &  \qquad \qquad + \frac{m(t_{Y_1}) + m(t_{Y_2})}{M(t_{Y_1}, {t_{Y2}})}\, \frac{p^{2}_{Y_1 Y_2}}{2\mu_{Y_1 Y_2}}
 \, +V^{S=-2}_{Y_1 Y_2}\nonumber \\[4pt]
& & \qquad\qquad  + \big(m(t_{Y_1}) + m({t_{Y_2}}) -  m_{\Xi} -  m_N \big),
\end{eqnarray}
    \begin{align} \label{eq:hamiltonian_transition}
\begin{split}
   H^{S=-2}_{Y_1 Y_2 ,\Xi N} = \sum_{i=1}^{A-1}   V^{S=-2}_{Y_{1} Y_{2} , \Xi i}\,,
\end{split}
\end{align}
\begin{eqnarray} \label{eq:hamiltonianXi}
 & & H_{\Xi}   = H^{S=0}_{\Xi}  \,   +  \,    H^{S=-2}_{\Xi} \nonumber\\
& & \qquad = \sum_{i < j=1}^{A-1} \Big( \frac{2p^{2}_{ij}}{M({\Xi})} \,+ V^{S=0}_{ij} \Big)\nonumber\\
 &  &\qquad  \,\,+ \sum_{i=1}^{A-1}
\Big( \frac{m_N + m_{\Xi}}{M({\Xi})} \,\frac{p^2_{\Xi i}}{2\mu_{\Xi i}} \,+ V^{S=-2}_{\Xi i }\Big),\\[2pt]
\nonumber
  \end{eqnarray}
 with    $|Y_1Y_2 \rangle= |\Lambda \Sigma\rangle$   or $|\Sigma\Sigma\rangle$.    
Here,  $m({t_{Y_1}}),  m(t_{Y_2})$ , $m_{\Xi}$ and $m_N$  are the  $Y_1$, $Y_2$, $\Xi$  hyperon   and nucleon rest masses, respectively.  $M$ denotes the total rest mass of the system, thus, 
 $M(t_{Y_1},t_{Y_2}) = m(t_{Y_1}) + m(t_{Y_2}) + (A-2)m_N$ and  $M(\Xi) = m_{\Xi} + (A-1)m_N$.   $\mu_{i \Xi}$ and $\mu_{Y_{1} Y_{2}}$ are the $\Xi$N  and YY reduced masses, respectively.   The  rest  mass differences  within the nucleon-  and  hyperon-isospin multiplets are neglected. $V^{S=0}$ and 
 $V^{S=-2}$ are the two-body NN and YY ($\Xi$N) potentials. Finally, 
the last term in Eq.~(\ref{eq:hamiltonian2Y}) accounts for
the difference in the  rest masses of   the hyperons  arising due to  particle conversions.

With the basis states defined in Eqs.~(\ref{eq:basisS=Xi},\ref{eq:basisS=Y1Y2}), the matrix elements of the Hamiltonian
Eqs.~(\ref{eq:hamiltonian2Y}-\ref{eq:hamiltonianXi}) can be evaluated analogously as done in \cite{Le:2021wwz}.  Likewise,
the final $\Xi$ wave functions and the corresponding  binding energies are directly obtained via the Lanczos eigenvalue iterations.

  \section{Results and discussion}\label{sec:results}
    %%%%%
        \begin{figure*}[htbp] 
      \begin{center}
      \hspace{0.3cm}{
           \subfigure[$E_{\mathcal{N}}(^{4}_{\Xi}\mathrm{H}(1^+,0)) $ as a function of $\omega$.]{\includegraphics[width=0.45\textwidth,trim={0.0cm 0.00cm 0.0cm 0 cm},clip]{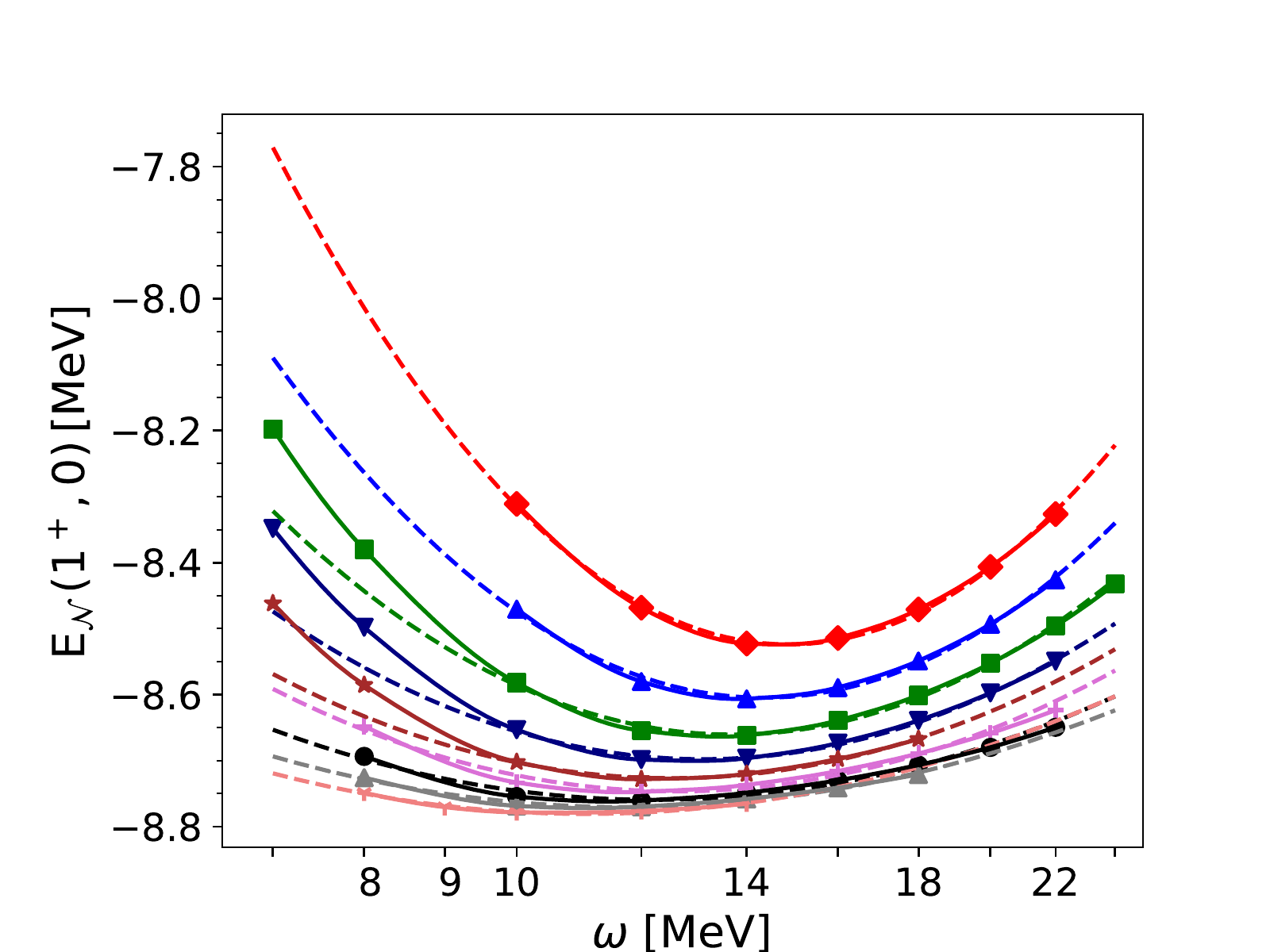}}
      \subfigure[$E( ^{4}_{\Xi}\mathrm{H}(1^+,0))$ as a function of $\mathcal{N}$.]{ \includegraphics[width=0.45\textwidth,trim={0.0cm 0.00cm 0.0cm 0 cm},clip]{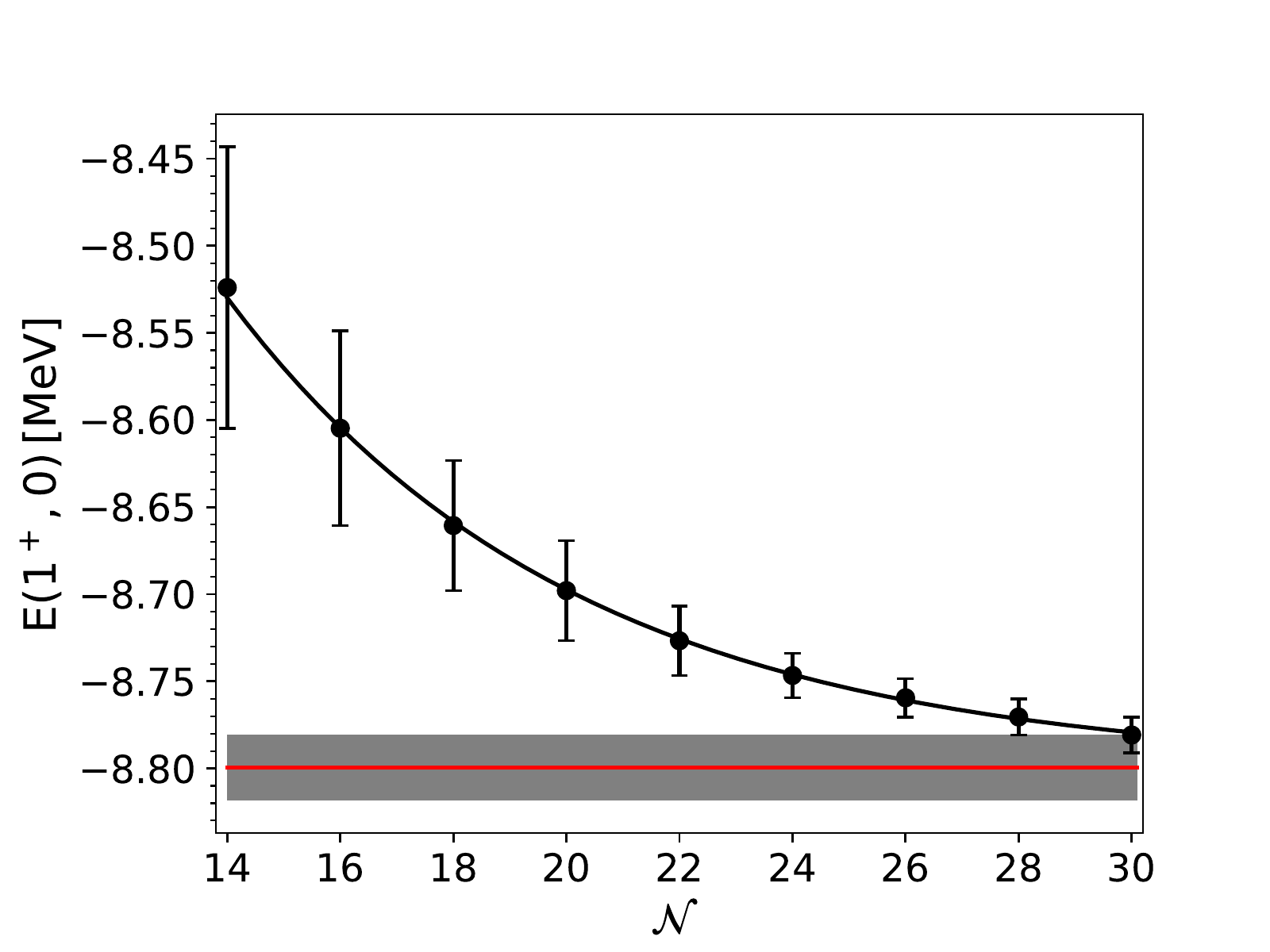}}}\\ 
         \vskip 0.15cm
      \hspace{0.3cm}{\subfigure[$B_{\Xi}(^{4}_{\Xi}\mathrm{H}(1^+,0))$ as a function of $\mathcal{N}$.]{\includegraphics[width=0.45\textwidth,trim={0.0cm 0.00cm 0.0cm 0.0cm},clip] {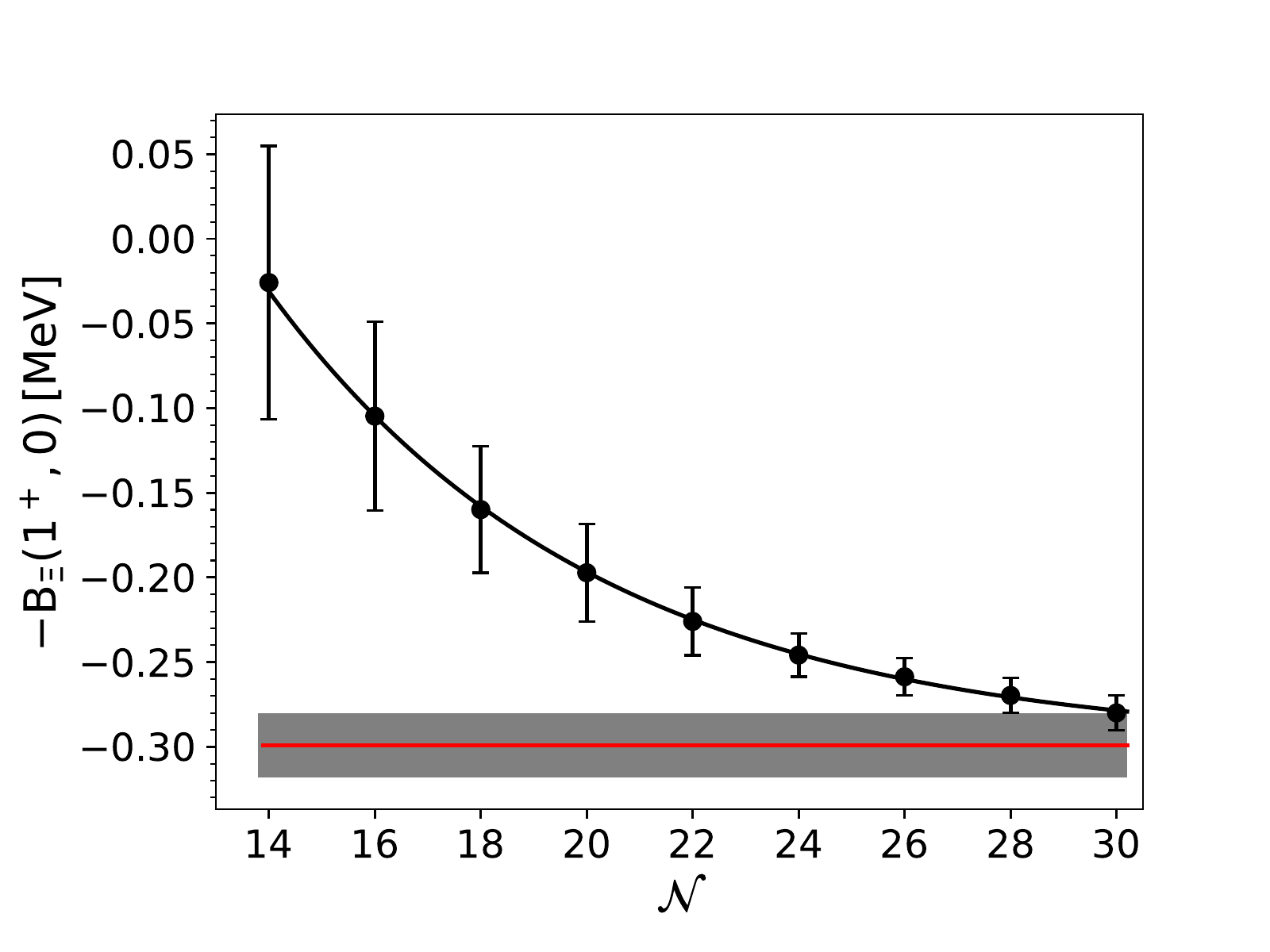}}
      \subfigure[$B_{\Xi}(^{4}_{\Xi}\mathrm{H}(1^+,0))$ as a function of $\lambda_{YY}$.]{ \includegraphics[width=0.45\textwidth,trim={0.0cm 0.00cm 0.0cm 0.0cm},clip]{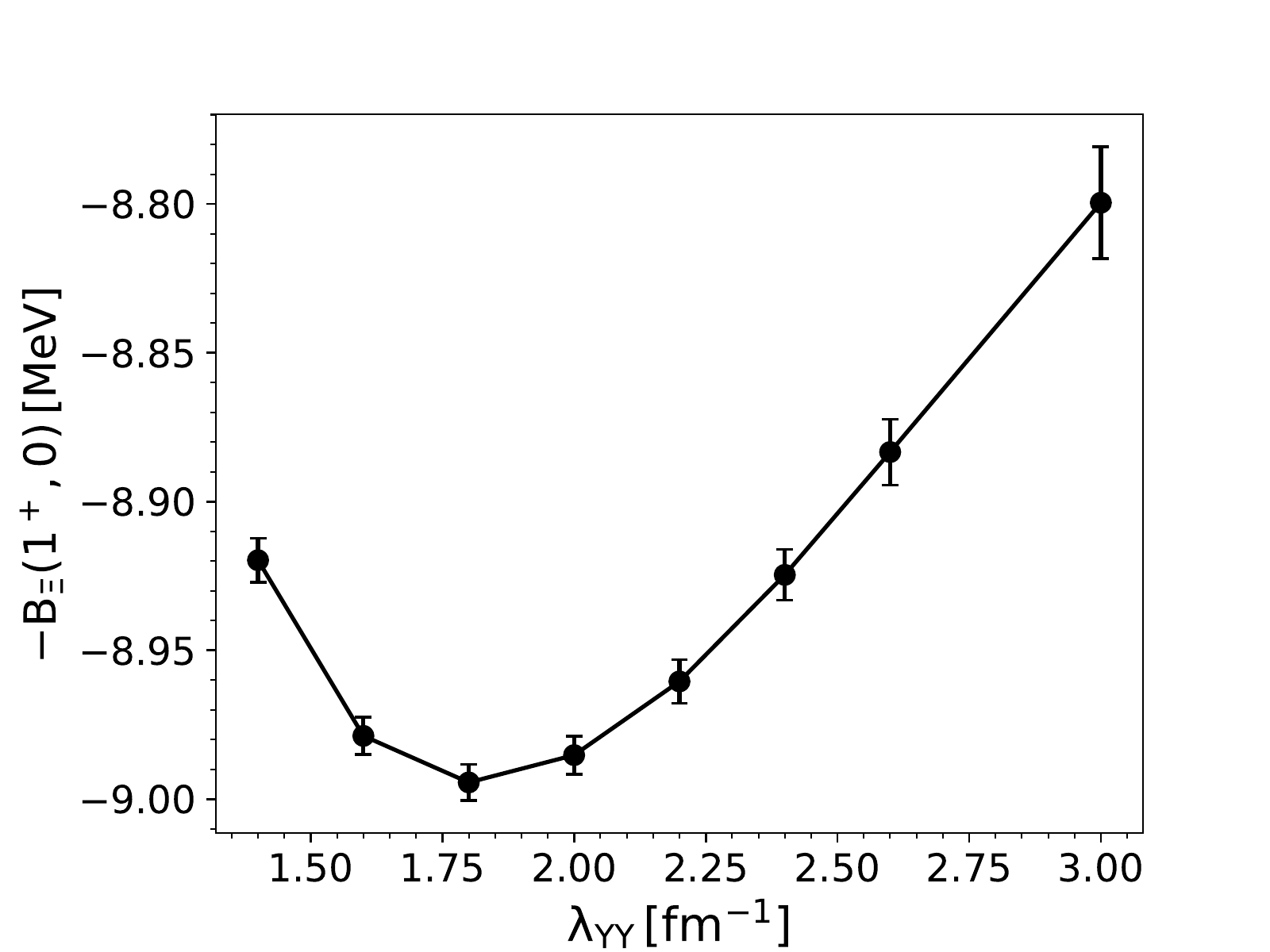}}}
      \end{center}
        \caption{ (a-c): binding energy $E$ and  $\Xi$ separation energy $B_{\Xi}$  for  $ ^{4}_{\Xi}\mathrm{H}(1^+,0)$ computed with   
        the YY-$\Xi$N interaction NLO(500), SRG-evolved to a flow parameter of $\lambda_{YY} =3.0 $ fm\textsuperscript{-1}. For
        the NN interaction the SMS N\textsuperscript{4}LO+(450) potential \cite{Reinert:2017usi} with $\lambda_{\mathrm{NN}}=1.6$~fm\textsuperscript{-1} is employed. $B_{\Xi}$ is measured with respect to the triton binding energy (which is $E(^3\mathrm{H}) =-8.5$~MeV for the 
        used NN interaction). 
        (a): Solid lines and symbols (with different colors)
        represent numerical results for different model spaces $\mathcal{N} =14 -30$, from top to bottom. 
        The dashed lines are obtained by using the ansatz Eq.\,(22) 
        in \cite{Le:2020zdu}. 
        (b-c): 
        Horizontal (red) lines with shaded areas  indicate the converged results and the corresponding uncertainties.
        (d): Dependence of $B_{\Xi}( ^{4}_{\Xi}\mathrm{H}(1^+,0))$  on the flow
        parameter $\lambda_{ YY }$.  
 }
    \label{fig:Convergence_4HXi_J=1-T=0}
         \end{figure*}
%%%%%%
  %\subsection{$ NNN \Xi$ }

  As it has been shown in \cite{Miyagawa:2021krh}, the $^3_{\Xi}\mathrm{H}$ hypernucleus is not bound with the 
  chiral YY-$\Xi$N NLO
  potential \cite{Haidenbauer:2018gvg}. Therefore, the lightest system that 
  we study here is NNN$\Xi$.  A $\Xi$ hyperon with
  isospin  $\frac{1}{2}$   can couple  to the core nucleus $^3\mathrm{H}/ ^3\mathrm{He}$ in its ground state $(\frac{1}{2}^+, \frac{1}{2})$   resulting in  several NNN$\Xi$ states with  $ (J^{\pi}, T)= (1^+,0)$, $(0^+,1)$, $(1^+,1)$ and $(0^+,0)$.  The first three states are found to be strongly bound in the  work by Hiyama \etal when the Nijmegen ESC08c potential is used. The HAL QCD potential, however, supports only
  one weakly bound NNN$\Xi$ state, namely the $(1^+,0)$  \cite{Hiyama:2019kpw}.  
  As discussed in Section \ref{sec:BBforces}, there are some differences in the predictions for the $\Xi$N phase shifts by the NLO(500) and HAL QCD 
  interactions. It is therefore interesting to calculate the $A=4$ system based 
  on the chiral potential in order to see whether such differences are
  manifest in the predictions for the NNN$\Xi$ binding energies.
  We also consider $^5_{\Xi}\mathrm{H}$. Due to the strong binding of the $\alpha$ particle core, the mass difference between $\Xi$N and $\Lambda \Lambda$ is partly removed \cite{Myint:1994qv} which makes this light hypernucleus 
  especially interesting. Likewise, theoretical predictions for 
  $^7_{\Xi} \mathrm{H}$ are of importance since this    system
  is expected to be investigated through the $^7{\mathrm{Li}} (K^-, K^+)$ reaction in upcoming experiments at J-PARC \cite{Fujioka:2021S}. Generally, 
  we expect that a consistent study of hypernuclei 
  for a range of mass numbers might provide constraints for the properties 
  of $\Xi$s in nuclear matter. In this work, we present such an 
  investigation for $\Xi$ hypernuclei up to $A=7$.
%%%%%

  In this exploratory 
  study, we do not take the $\Xi^-$-$\Xi^0$ mass difference of 6.85~MeV into 
  account and assume isospin symmetry by assigning to each state a definite isospin. 
  We believe that this is justified to identify states that are possibly bound
  but stress that the impact of the mass difference and possible 
  isospin breaking by other contributions like the Coulomb interaction 
  should be analysed in a future study. For the $A=3$ and $A=5$ $\Xi$ hypernuclei, the isospins  are  well defined since the corresponding core nucleus is predominantly in its  isospin zero state. 
   For $A=7$, we follow the choice of Hiyama \etal \cite{Hiyama:2019kpw} and only consider the $(1/2^+,\, 3/2)$ 
  state. For this state, it is natural to expect that the $\Xi^-$-$^6 \mathrm{He}(0^+,1)$ component is the dominant one. 
 The situation is less clear for 
  $A=4$ since, {\it a priori}, none of the possible isospins is 
  favored by the 3N core. Below we have therefore assumed isospin symmetry
  and give separate results for $T=0$ and $T=1$ states.
  Our results for the contribution of the Coulomb interaction suggest that 
  this is a reasonable approximation since the binding is still predominately  
  due to the strong interaction. However, a more careful analysis also taking the 
  cascade mass difference into account should be performed in the future.

 As mentioned earlier, in order to eliminate the effect of finite-basis truncation on the binding energies, we follow the two-step extrapolation procedure as explained in  \cite{Le:2020zdu}.
 The $\omega$- and $\mathcal{N}$-space extrapolations for the binding energy of the $^4 _{\Xi}{\rm H}(1^+,0)$ state, $E(1^+,0)$, are illustrated
in panels (a) and (b) of Fig.~\ref{fig:Convergence_4HXi_J=1-T=0}, respectively.    To obtain  the  converged $\Xi$ separation energy $B_{\Xi}$, we perform an analogous  exponential fit  on the quantities
  $B_{\Xi}(\mathcal{N}) = E_{\mathcal{N}}(1^+,0)  - E_{\mathcal{N}}(^3\mathrm{H})$, see also panel (c).  Here  $E_{\mathcal{N}}(1^+,0) $,
  $ E_{\mathcal{N}}(^3\mathrm{H})$ are  the hypernuclear and nuclear binding energies, respectively, obtained at  their
  optimal HO frequencies for a given model space $\mathcal{N}$. For the separation energies, we cannot expect a monotonic convergences {\it a priori}. But for 
  the results shown here, we observed that this is the case and that 
  an exponential fit is appropriate. 
  The same approach was used for the separation energies 
  obtained in our earlier studies \cite{Le:2020zdu,Le:2021wwz}.   Note that
  the error bars shown in panels (b)-(c) are given
by the difference to the next model space.  These error bars  are not meant to provide a realistic uncertainty estimate, but rather to assign 
 relative weights for  the following  extrapolation to $\mathcal{N} \rightarrow \infty$.  Clearly, well-converged results for both   $E(1^+,0)$ and  $B_{\Xi}(1^+,0)$ are achieved  for  model spaces up to $\mathcal{N}_{max}=30$. For the NN interaction alone, the triton energy calculation is, however, converged already for model spaces  $\mathcal{N}_{max} =24$.   Similar convergence patterns   are  also observed for  the binding (separation) energies  of the other  states in NNN$\Xi$  
  and of  the  $^5_{\Xi}\mathrm{H}(\frac{1}{2}^+,\frac{1}{2})$,  $^7_{\Xi}\mathrm{H}(\frac{1}{2}^+,\frac{3}{2})$
 hypernuclei. The convergence for $A=5$ and $A=7$
 is generally faster since the separation energies are larger. Therefore, our current limits of $\mathcal{N}_{max} =16$ and 12 for $A=5$ and 7, respectively, still allow an accurate 
 determination of the energies as can be seen from Table~\ref{tab: A=4-7system}.
 
    In order to minimize the contribution of the Coulomb interaction, we use the 
    $^4 _{\Xi}{\rm n}$ states for isospin $T=1$. As can be seen in Table~\ref{tab: A=4-7system}, both of these states,  $^4_{\Xi}{\rm n}(0^+,1)$ and $^4_{\Xi}\rm{n}(1^+,1)$,  are clearly bound for the chiral interaction. 
    It turns out that  their 
    binding energies are comparable to that one for $^4 _{\Xi}{\rm H}(1^+,0)$. 
    The $^4 _{\Xi}{\rm H}(0^+,0)$ state, on the other hand,  is unbound for the chiral interactions 
    at NLO. 

  The converged  $B_{\Xi}(1^+,0)$ computed for a wide range of SRG-flow parameter,
  $ 1.4 \le \lambda_{ YY}  \le 3.0$ fm\textsuperscript{-1}, are presented in     panel (d) of Fig.~\ref{fig:Convergence_4HXi_J=1-T=0}.
  One sees that the overall variation of $B_{\Xi}(1^+,0)$  is visible, about $190 \pm 30$ keV.  It is also larger than the dependence of $B_{\Lambda\Lambda}$ on the SRG-flow parameter, which was found to be of the order of 100 keV \cite{Le:2020zdu}.  This
  may be related to the fact that, unlike for the $\Lambda\Lambda$ case 
  where the coupling to the pion is suppressed by isospin conservation, 
  $\pi$ exchange contributes to the $\Xi$N interaction at leading order. 
  Long-range interactions are likely to be more strongly affected by the
  SRG evolution. However, the variation is much smaller than 
  the one observed for single $\Lambda$ hypernuclei (see e.g. \cite{Le:2020zdu}).
  For the $A=5$ and 7 $\Xi$ hypernuclei, we observed 
  similarly large absolute variations, but still they are relatively smaller as compared to the estimated $\Xi$ separation energies. Therefore, in all cases, 
  the SRG dependence is small enough that it does not affect 
  conclusions on the existence of bound states. 
 Therefore, in the  following discussion, we will present results for a specific flow parameter, namely 
 $\lambda_{YY} = 1.6$~fm\textsuperscript{-1}. 

  %%%%%
   \begin{table}
    \renewcommand{\arraystretch}{2.0}
 \vskip 1 cm
\begin{center}
   \setlength{\tabcolsep}{0.17cm}
\begin{tabular}{|c|c c |}
\cline{1-3}
   &  $B_{\Xi}  $ [MeV]  &  $ \Gamma$   [MeV]   \\ \hhline{  = = =}
$ ^4_{\Xi}\mathrm{H}  (1^+,0) $   &  $0.48 \pm 0.01 $ &   0.74\\
%\cline{1-3}
$ ^4_{\Xi}\mathrm{n}  (0^+,1) $ &  $0.71 \pm 0.08$ & 0.2\\
%\cline{1-3}
$ ^4_{\Xi}\mathrm{n} (1^+,1) $    & $ 0.64 \pm 0.11$ & 0.01\\ 
%\cline{1-3}
$ ^4_{\Xi}\mathrm{H} (0^+,0) $    & - &  - \\ \hhline{  = = =}
$^5_{\Xi}\mathrm{H}(\frac{1}{2}^+,\frac{1}{2})$ &   $2.16 \pm 0.10$ &  0.19  \\ \hhline{  = = =}
$^7_{\Xi}\mathrm{H}(\frac{1}{2}^+,\frac{3}{2})$ &   $3.50 \pm 0.39$ &  0.2  \\ 
  \cline{1-3}
\end{tabular}
\end{center}
\caption{  $\Xi$ separation energies $B_{\Xi}$ and  estimated decay widths $\Gamma$ for $A=4-7$ $\Xi$ hypernuclei.   All calculations are based on  the YY-$\Xi$N interaction NLO(500) and the NN interaction SMS 
N\textsuperscript{4}LO+(450). Both potentials are SRG-evolved to a 
flow parameter of $\lambda_{\mathrm{NN}} = \lambda_{YY} 
=1.6$~fm\textsuperscript{-1}. The values of   
$B_{\Xi}$ in NNN$\Xi$, $^5_{\Xi}\mathrm{H}$ and $^7_{\Xi}\mathrm{H}$
  are  measured with respect to the binding energies of the core nuclei $^3\mathrm{H}$, $^4{\mathrm{He}}$ and $^6{\mathrm{He}}$, respectively.  }
\label{tab: A=4-7system}
\end{table}
%%%%%%

 The predicted separation energies $B_{\Xi}$ for the  $A=4-7$ $\Xi$ hypernuclei  are listed in Table~\ref{tab: A=4-7system}. 
 We verified that all the bound states established here  are  predominantly due to the strong $\Xi$N interaction. 
   The $\Xi^- p$ Coulomb interaction contributes   roughly  200, 600, and  400 keV to the  binding energies of  NNN$\Xi$, $^5_{\Xi}\mathrm{H}$ and  $^7_{\Xi}\mathrm{H}$, respectively.
 Table~\ref{tab: A=4-7system} provides also an estimate of
 the corresponding decay width $\Gamma$. These widths have
 been evaluated perturbatively by adapting the procedure followed
 by Hiyama \etal   \cite{Hiyama:2008fq,Hiyama:2019kpw}.  Hiyama \etal have used
 the imaginary part of the $G$ matrix.  Here, we employ 
 the $\Xi$N $T$-matrix in the $^{11}S_0$ 
 state from the original potential that includes 
 the $\Xi$N-$\Lambda\Lambda$ coupling \cite{Haidenbauer:2018gvg} instead.
 Schematically the width amounts to \break
 $\Gamma \simeq -2\, {\rm Im}\, 
 \langle \Psi_{B_\Xi} |T_{\Xi N - \Xi N}|\Psi_{B_\Xi}\rangle$
 and involves the pertinent hypernuclear wave function $\Psi_{B_\Xi}$ 
 and the (off-shell) $\Xi$N $T$-matrix 
 at the sub-threshold energy corresponding to the bound state.
 One can clearly see that the three states $(1^+,0)$, $(0^+,1)$ and $(1^+,1)$ in NNN$\Xi$ are only weakly bound, possessing quite 
 similar $B_{\Xi}$'s but substantially different decay widths.    
 Interestingly, our result for $B_{\Xi}({\rm NNN}\Xi(1^+,0))$ is close to
    that for the HAL QCD potential, reported in \cite{Hiyama:2019kpw}, 
    although the $(0^+,1)$ and $(1^+,1)$ states are unbound
    for the HAL QCD interaction. There are 
    substantial (but not surprising) differences between our 
    separation energies \break $B_{\Xi}({\rm NNN}\Xi)$ and the predictions
    \cite{Hiyama:2019kpw} for the ESC08c potential \cite{Nagels:2015dia}. 
    According to the discussion in Ref.~\cite{Hiyama:2019kpw}, 
    it is the strong attraction in the $^{33}S_1$ and $^{13}S_1$ 
    channels that is responsible for the rather large binding energies
    predicted for that $\Xi$N model in the $(1^+,0)$ and $(1^+,1)$ 
    states. 
    
    In \ref{app:pwestimate}, we summarize the relative 
    weights of the different partial wave channels to the effective $\Xi$N
    interaction in the $s$-shell $\Xi$ hypernuclei. Although
    such an estimate is rather rough, it can nevertheless
     help to understand the pattern of different bound states found.

    Our results for the $^5_{\Xi}\mathrm{H}$ separation energy and
    decay width are $B_{\Xi}(^5_{\Xi}\mathrm{H})=2.16 \pm 0.1$~MeV 
    and $\Gamma(^5_{\Xi}\mathrm{H})=0.19$  MeV, respectively. 
    Oddly enough, these values agree roughly with the estimations by 
    Myint and Akaishi \cite{Myint:1994qv} of $1.7$~MeV and $0.2$~MeV, 
    respectively. However, given the differences in the underlying 
    interactions and specifically in the employed approaches, this is 
    certainly accidental. We further note that in contrast 
    to our finding where $^5_{\Xi}\mathrm{H}$ is bound primarily due to the strong $\Xi$N interaction, the authors in \cite{Myint:1994qv} state that the binding energy of $1.7$~MeV in $^5_{\Xi}\mathrm{H}$ largely comes from the  $^4\mathrm{He}$-$\Xi^-$ Coulomb  interaction. The mechanism for the narrow width of  $^5_{\Xi}\mathrm{H}$ has been investigated in \cite{Dover:1994qs,Kumagai-Fuse:1995zeb}.   
    Recently, Friedman and Gal, employing an optical potential, also reported a quite 
    similar result for 
    $^5_{\Xi}\mathrm{H}$ ($B_{\Xi}(^5_{\Xi}\mathrm{H})=$ \\ 2.0~MeV)  \cite{Friedman:2021rhu}. 
    But also here the agreement seems to be more or less accidental 
    given that the
    $\Xi$-nuclear interaction used as starting point in that work is with 
    $U_\Xi \lesssim -20$~MeV significantly more attractive than the one predicted by the chiral $\Xi$N potential employed in the present
    study which is only around $U_\Xi \approx -9$~MeV \cite{Kohno:2019oyw} 
    as mentioned above. 
    
 The prediction of the chiral $\Xi$N interaction for $^7_{\Xi}\mathrm{H}$ $(\frac{1}{2}^+, \frac{3}{2})$, $B_{\Xi}(^7_{\Xi}\mathrm{H})=3.50 \pm 0.39$~MeV,
 is only slightly larger than the binding energy of 3.15~MeV reported by Fujioka \etal 
 \cite{Fujioka:2021S,Fujioka:2021T} for the HAL QCD interaction,
 based on a calculation within a four-body ($\alpha nn \Xi$) cluster model
 \cite{Hiyama:2008fq,Hiyama:2021P}.
 An earlier study utilizing older $S=-2$ potentials from the Nijmegen group 
 indicated somewhat 
 smaller binding energies \cite{Ajimura:2019,Hiyama:2008fq}. 
   Finally, as one can see from Table~\ref{tab: A=4-7system},  the $^7_{\Xi}\mathrm{H}\,(\frac{1}{2}^+, \frac{3}{2})$ state is also very narrow, with a width of $\Gamma=0.2$~MeV.
     
%%%%%%%%%
 \begin{table}
    \renewcommand{\arraystretch}{2.0}
 \vskip 1 cm
\begin{center}
   \setlength{\tabcolsep}{0.074cm}
\begin{tabular}{|c|ccc c c |c|}
\cline{1-7}
   & \multicolumn{5}{c|}{ $V^{S=-2} $}      &\multicolumn{1}{c|}{ E} \\
   &  $^{11}S_{0}$ & $^{31}S_{0}$  &  $^{13}S_{1}$  & $^{33}S_{1}$   & total &      \\ \hhline{ = = = = = = =}
$ ^4_{\Xi}\mathrm{H} (1^+,0) $  &  -1.95  & 0.02  & -0.7   & -2.31    & -5.21& -8.97\\
%\cline{1-7}
$ ^4_{\Xi}\mathrm{n} (0^+,1) $  &  -0.6  & 0.25  & -0.004   & -0.74  & -1.37 & -9.07\\
%\cline{1-7}
$ ^4_{\Xi}\mathrm{n} (1^+,1) $  &  -0.02 & 0.16  & -0.13   & -1.14  & -1.30 &  -9.0\\
%\cline{1-7}
$ ^4_{\Xi}\mathrm{H}  (0^+,0) $  &  -0.002 & 0.08  & -0.01   & -0.006  & -0.11 &  -6.94\\ \hhline{ = = = = = = = }
$ ^5_{\Xi }\mathrm{H} (1/2^+,1/2) $  &  -0.96 & 0.94  & -0.58   & -3.63  & -4.88 &  -31.43\\ \hhline{= = = = = = = }
$ ^7_{\Xi}\mathrm{H} (1/2^+,3/2) $  &  -1.23 & 1.79  & -0.79   & -6.74  & -8.04 & -33.22 \\ 
\cline{1-7}
\end{tabular}
\end{center}
\caption{Contributions of different  partial waves to $ \langle  V^{S=-2} \rangle$ (first five columns),  and the total binding energy (last column) for the $A=4-7$  $\Xi $ hypernuclei. The results are extracted at  $\mathcal{N}=28$, 
$\omega =10$ MeV for NNN$\Xi$,  at   $\mathcal{N} =14 $,   $\omega =16$ MeV for $^5_{\Xi }\mathrm{H} $ and at 
$\mathcal{N}=10$, $\omega=16$~MeV for $^7_{\Xi}\mathrm{H}$.    All energies are  given in MeV. Same interactions as in Table~\ref{tab: A=4-7system}.  Note that the calculated binding energy of $^3\mathrm{He} (^3\mathrm{H} )$  is $-7.79\, (-8.50)$~MeV. }
\label{tab:expectationVYY}
\end{table}
%%%%%%
To shed light on the relation between the properties of the 
chiral $\Xi$N potential 
and the binding of the $A=4-7$ $\Xi$ systems, we provide
in Table~\ref{tab:expectationVYY} the contributions of different $\Xi$N 
partial waves to the expectation value of the $S=-2$ potential 
$\langle V^{S=-2}\rangle$. These results are computed 
at  $\mathcal{N} =28, \omega=10$ MeV for NNN$\Xi$,
at $\mathcal{N}=14, \omega= 16$~MeV for $^5_{\Xi} \mathrm{H}$ and  at $\mathcal{N}=10, \omega= 16$~MeV for $^7_{\Xi} \mathrm{H}$.  Here the second largest model space is chosen  for each system in order to save computational resources. And, $\omega$ is the corresponding  optimal  HO frequency for  the chosen model space. For completeness, the  energy expectation values  are also shown in the last column of  Table~\ref{tab:expectationVYY}. Clearly,  in all the considered states except NNN$\Xi(0^+,0)$  the attractive $\Xi$N interaction in the $^{33}S_{1}$ channel plays the most important role in binding the systems. 
It accounts for more than $50\%$ of the expectation value 
$\langle V^{s=-2} \rangle$. 
While the attraction in the  $^{11}S_{0}$ channel is essential as well for  NNN$\Xi(1^+,0)$ and $(0^+,1)$ (amounting to more than $30 \%$ of $\langle V^{S=-2}\rangle$), its contribution becomes less significant  in other states.  Furthermore, the  $\Xi N$  repulsion in $^{31}S_{0}$ contributes predominantly to the expectation value $\langle V^{S=-2}\rangle$ of NNN$\Xi(0^+,0)$ 
(naturally with opposite sign), which causes the system to be unbound. 
The expectation 
value $\langle V^{S=-2}(^{31}S_{0})\rangle$ is also
sizable for $^5_{\Xi}\mathrm{H}$ and $^7_{\Xi}\mathrm{H}$, however, 
its effect is largely canceled by the  attraction in  the $^{11}S_{0}$ channel.  
 
%%%%%
   \begin{table}
    \renewcommand{\arraystretch}{2.0}
 \vskip 1 cm
\begin{center}
   \setlength{\tabcolsep}{0.085cm}
\begin{tabular}{|c|ccc c c |}
\cline{1-6}
   & \multicolumn{5}{c|}{ $|\Xi N\rangle $}    \\
   &  $|^{11}S_{0}\rangle$ & $|^{31}S_{0}\rangle$  &  $|^{13}S_{1}\rangle$  & $|^{33}S_{1}\rangle$   & $J \ge 2$      \\ \hhline{  = = = = = =}
$ ^4_{\Xi}\mathrm{H} (1^+,0) $  &  12.88  & 0.18  & 25.91   & 35.72    & 24.80 \\
%\cline{1-6}
$ ^4_{\Xi}\mathrm{n}  (0^+,1) $  &  8.24  & 13.32  & 0.23   & 23.29  & 54.73 \\
%\cline{1-6}
$ ^4_{\Xi}\mathrm{n}  (1^+,1) $  &  0.14 & 9.22  & 9.83   & 33.08 & 47.56 \\
%\cline{1-6}
$ ^4_{\Xi}\mathrm{H}  (0^+,0) $  &  0.02 & 11.87  & 14.65   & 0.11  & 73.33 \\ \hhline{ = = = = = =  }
$ ^5_{\Xi }\mathrm{H} (1/2^+,1/2) $  &  4.82 & 12.18  & 14.37   & 35.53  & 32.59 \\ \hhline{ = = = = = = }
$ ^7_{\Xi}\mathrm{H} (1/2^+,3/2) $  &  3.71 & 12.92  & 11.11   & 38.36 & 32.94 \\ 
\cline{1-6}
\end{tabular}
\end{center}
\caption{ Probabilities (in \%) of finding a $\Xi$N pair  in different partial-wave states in the wave functions of  $A=4-7$ $\Xi$ hypernuclei. Same interactions and model spaces as in Table~\ref{tab: A=4-7system}. Note that for each system all probabilities sum up to the probability of finding a $\Xi$ hyperon in that system.  }
\label{tab:ProbXiN_indiff_partialwave}
\end{table}
%%%%%
Complementary to Table~\ref{tab:expectationVYY},  the binding of 
the $A=4-7$ hypernuclei 
can also be understood from Table~\ref{tab:ProbXiN_indiff_partialwave}, where probabilities of finding a $\Xi$N pair, $P_{\Xi \mathrm{N}}$,  in   different partial-wave states  are listed.  One clearly notices that,  
in most of the systems,
 a $\Xi$N pair is predominantly found in those channels with $J \leq 1$ and in particular  in  the  $^{33}S_{1}$, except for  the unbound $^4_{\Xi}\mathrm{H}(0^+,0)$ state.  In addition, the two extremely small probabilities  $P_{\Xi \mathrm{N}}(^{11}S_{0}) =0.0 2 \%$ and  $P_{\Xi N}(^{33}S_{1})=0.11 \%$  in $^4_{\Xi}\mathrm{H}(0^+,0)$ are obvious manifestations 
 of the small expectation values $V^{S=-2}(^{11}S_{0}) =-0.002$~MeV
 and \break $V^{S=-2}(^{33}S_{1}) =-0.006$~MeV  listed  in Table.~\ref{tab:expectationVYY}.  Furthermore,  the strong variation of 
 $P_{\Xi \mathrm{N}}(^{11}S_{0}) $ in different states of the $A=4-7$  hypernuclei  clearly explains the large difference in the decay widths estimated for these systems, see Table~\ref{tab: A=4-7system}.

As discussed in Section~\ref{sec:BBforces}, we had to omit 
the $\Lambda\Lambda-\Xi$N coupling in the J-NCSM application 
and we compensated that by a small modification of the 
$\Xi$N potential strength in the $^{11}S_0$ state.   
It is reassuring to see that the overall effect of this 
partial wave on the binding energies is not too large. 
Specifically, the existence of the predicted bound states 
does not depend on its precise contribution, as 
can be read off from
Tables~\ref{tab:expectationVYY} and \ref{tab:ProbXiN_indiff_partialwave}.
In fact, the slightly more attractive $^{11}S_0$ interaction
predicted by the original $\Xi$N potential,
see Fig.~\ref{fig:XiNphase_shift}, implies that all found $\Xi$ 
hypernuclei could be simply minimally more bound. 

 %%%
  \section{Conclusions}\label{sec:Concl}
  In this work, we employed the Jacobi NCSM in combination with the 
  chiral NLO(500) $\Xi $N potential to explore $A=4-7$ $\Xi$ 
  hypernuclei. 
  Particle conversions like $\Lambda\Sigma -\Xi$N$ -  \Sigma\Sigma $ 
  are fully taken into account, while the transition
  $\Lambda\Lambda - \Xi $N  is omitted and its contribution is incorporated effectively by re-adjusting the 
  strength of the $V_{\Xi N}$ potential appropriately. 
  The latter approach facilitates a proper convergence 
  of the energy calculations to the lowest lying $\Xi$ states. Furthermore, to speed up the convergence, the $\Xi$N potential is SRG-evolved to a wide range of flow parameters. The effect of SRG evolution on the $\Xi$ separation energies is in general small, 
  but, it is slightly larger than that observed for
  $\Lambda\Lambda$ hypernuclei. We found three loosely bound states $(1^+,0)$, $(0^+,1)$  and $(1^+,1)$ for the NNN$\Xi$ system and 
   more tightly bound $^5_{\Xi}\mathrm{H}$, $^7_{\Xi}\mathrm{H}$ hypernuclei. These $\Xi$ systems are bound 
  predominantly due to the attraction of the chiral $\Xi$N potential
  in the $^{33}S_{1}$ channel. On the other hand, the
  repulsive nature in $^{31}S_{0}$ prevents the binding of the  NNN$\Xi(0^+,0)$ state. All the investigated $\Xi$ bound states are 
  predicted to have very small decay widths. 
  
  In view of these results, which are based on an interaction that is fully consistent with presently available experimental constraints,
  and well in line with current lattice QCD results \cite{HALQCD:2019wsz}, 
  it seems likely that light $\Xi$ hypernuclei exist. 
  Experimental confirmation is certainly challenging. However,
  theoretical estimates for yields of $A=4$ hypernuclei   \cite{Steinheimer:2012tb} as well as actual measurements 
  of $^4_{\Lambda}\mathrm{H}$, $^4_{\Lambda}\mathrm{He}$
  by the STAR Collaboration \cite{Leung:2021} raise hopes
  that NNN$\Xi$ bound states can be detected in heavy ion
  collisions in the not too far future. Also a bound 
  $^7_{\Xi}\mathrm{H}$ system could be produced and 
  studied in the $^7$Li$(K^-,K^+)$ reaction, cf. the proposal
  P75 for J-PARC \cite{Ajimura:2019}. 
  Once these new experimental results are available, 
  they will provide new insights into the properties of $S=-2$ BB interactions. The current manuscript sets up a framework that allows one to exploit these insights to constrain BB interactions in the future. 
  
\vskip 0.3cm
{\bf Acknowledgements:} This work is supported in part by the NSFC and the Deutsche Forschungsgemeinschaft (DFG, German Research
Foundation) through the funds provided to the Sino-German Collaborative
Research Center TRR110 ``Symmetries and the Emergence of Structure in QCD''
(NSFC Grant \break No. 12070131001, DFG Project-ID 196253076 - TRR 110). We
also acknowledge support of the THEIA net-working
activity of the Strong 2020 Project. The numerical calculations
have been performed on JURECA and the
JURECA booster of the JSC, J\"ulich, Germany. The
work of UGM was supported in part by the Chinese
Academy of Sciences (CAS) President's International
Fellowship Initiative (PIFI) (Grant No. 2018DM0034)
and by VolkswagenStiftung (Grant No. 93562).

\appendix

\section{Estimate of partial wave contributions}
\label{app:pwestimate}

In this appendix, we summarize approximate partial wave contributions 
to $s$-shell $\Xi$ hypernuclei. 
The relations are similarly derived as the ones for $\Lambda$ hypernuclei and the $\Lambda N$ potential \cite{Gibson:1994yp}. For these rough 
estimates it is assumed that there is no
particle conversion contributing. Additionally, the $\Xi$ hypernucleus exhibits a clear core-$\Xi$ structure. Both, the core nucleons and the $\Xi$ are in $s$-wave 
states.  To justify these assumptions,  we provide in Tables~\ref{tab:Probcore_indiff_partialwave_a4}
and \ref{tab:Probcore_indiff_partialwave_a57} probabilities of finding the nucleons in certain 
angular momentum and isospin states and together with $\Xi^-$ or $\Xi^0$. For the isospin zero states of $A=4$, the hypernucleus seems to be dominated by the $^3$He/$^3$H component together with 
$\Xi^-$ and $\Xi^0$, respectively. The choice of $T=0$ enforces that 
both parts contribute equally. For the other 
isospin and $A=5$  and 7, the total charge of the systems is chosen such that the $\Xi^-$ 
contribution dominates in conjunction with the expected core nuclei. 
In the approximation that the hypernucleus only contains these dominant components,
one obtains for the effective interactions 
%%%%
  \begin{eqnarray}  \label{eq:3HXi1}
&  ^3_{\Xi} \mathrm{H} (\frac{1}{2}^+, \frac{1}{2}): \tilde{V}_{\Xi N }   &  \approx \frac{3}{16}  V_{\Xi N }^{^{11}S_{0}}+
 \frac{9}{16}  V_{\Xi N }^{^{31}S_{0}}
\,+\,  \frac{1}{16}  V_{\Xi N }^{^{13}S_{1}}    \nonumber \\[3pt]
&   & \quad +\,
 \frac{3}{16}  V_{\Xi N }^{^{33}S_{1}}
  \end{eqnarray}
  
    \begin{eqnarray}  \label{eq:3HXi2}
  &  ^3_{\Xi} \mathrm{H} (\frac{3}{2}^+, \frac{1}{2}): \tilde{V}_{\Xi N }  &  \approx \frac{1}{4}  V_{\Xi N }^{^{13}S_{1}} +  \frac{3}{4}  V_{\Xi N }^{^{33}S_{1}}
      \end{eqnarray}
      
        \begin{eqnarray}  \label{eq:4HXi10}       
    ^4_{\Xi}\mathrm{H}(1^+,0):  \tilde{ V}_{\Xi N}  \approx \frac{1}{6} V^{^{11}S_0}_{\Xi N} \,+ \,
\frac{1}{3} V^{^{13}S_1}_{\Xi N }   \,  +\,\frac{1}{2}   V^{^{33}S_1}_{\Xi N}
  \end{eqnarray}
      
             \begin{eqnarray}  \label{eq:4HXi01}   
   & ^4_{\Xi}\mathrm{H}(0^+,1): \tilde{V}_{\Xi N} & \approx \frac{1}{6} V^{^{11}S_0}_{\Xi N } \,+ \,
\frac{1}{3} V^{^{31}S_0}_{\Xi N } 
\,  + \,\frac{1}{2} V^{^{33}S_1}_{\Xi N } 
  \end{eqnarray}      

      \begin{eqnarray}  \label{eq:4HXi11}    
    & ^4_{\Xi}\mathrm{H}(1^+,1):  \tilde{ V}_{\Xi N } &  \approx \frac{1}{6} V^{^{31}S_0}_{\Xi N } \,+ \,
\frac{1}{6} V^{^{13}S_1}_{\Xi N } 
\,  +\,   \frac{2}{3} V^{^{33}S_1}_{\Xi N} 
    \end{eqnarray}
    
         \begin{eqnarray}  \label{eq:4HXi00}       
   ^4_{\Xi}\mathrm{H}(0^+,0):    \tilde{V}_{\Xi N } \approx   \frac{1}{2}V^{^{31}S_0}_{\Xi N}  \,+ \,
 \frac{1}{2} V^{^{13}S_1}_{\Xi N}
   \end{eqnarray}
   
  \begin{eqnarray}  \label{eq:5HXi}     
&  ^5_{\Xi}\mathrm{H}(\frac{1}{2}^+,\frac{1}{2}): \tilde{V}_{\Xi N } & \approx \frac{1}{16} V^{^{11}S_0}_{\Xi N} \,+ \,
\frac{3}{16} V^{^{31}S_0}_{\Xi N } \,+\, \frac{3}{16} V^{^{13}S_1}_{\Xi N }  \nonumber \\[5pt]
&  & \quad  + \,    \frac{9}{16} V^{^{33}S_1}_{\Xi N }.\\ \nonumber
  \end{eqnarray} 
 %%%%%
 It clearly follows  from Eqs.~(\ref{eq:3HXi1}-\ref{eq:5HXi}) that the $^{33}S_{1}$ $\Xi$N interaction strength dominates the $(\frac{3}{2}^+,\frac{1}{2})$ state in  $^{3}_{\Xi}\mathrm{H}$,  the $(1^+,0)$, $(0^+,1)$ and $(1^+,1)$ states in NNN$\Xi$ and the $^{5}_{\Xi}\mathrm{H}$ hypernucleus.  The repulsive $^{31}S_{0}$ and weakly attractive $^{13}S_{1}$ potentials contribute practically with equal weight to  NNN$\Xi\,(0^+,0)$,  whereas  the contribution from the $^{31}S_{0}$  channel  is dominant in   $^{3}_{\Xi}\mathrm{H}(\frac{1}{2}^+,\frac{1}{2})$.

   \begin{table*}
    \renewcommand{\arraystretch}{2.0}
 \vskip 1 cm
\begin{center}
   \setlength{\tabcolsep}{0.085cm}
\begin{tabular}{|c|c |c|c| c| }
\cline{1-5}
$(J_{core},T_{core}, m^t_{\Xi})$ &  $^4_{\Xi}H(1^+,0)$  &   $^4_{\Xi}n(0^+,1)$  & $^4_{\Xi}n(1^+,1)$ &
$^4_{\Xi}\mathrm{H}(0^+,0)$    
        \\ \hhline{  = = = = = }
$ (\frac{1}{2}, \frac{1}{2}, -\frac{1}{2})$  &   49.66   & 97.48  & 97.44  & 49.98  \\ \hhline{  - - - - - }
$ (\frac{1}{2}, \frac{1}{2}, \frac{1}{2})$  &   49.66   &  --   & --  & 49.98  \\ \hhline{  - - - - - }
$ (\frac{1}{2}, \frac{3}{2}, \frac{1}{2})$  &   --   &  0.54   & 0.55 & --  \\  \hhline{  - - - - - }
$ (\frac{1}{2}, \frac{3}{2}, -\frac{1}{2})$  &   --   &  1.6   & 1.6  & -- \\ \hhline{  = = = = = }
$ \text{others}$  &   0.16   &  0.17   &  0.22  &  0.02    \\ 
\cline{1-5}
\end{tabular}
\end{center}
\caption{ Probabilities (in \%) of finding the nucleons in a total $J_{core}$ and $T_{core}$ 
angular momentum and isospin state and with a $\Xi$ hyperon with third component of isospin $m^t_\Xi$ 
 in the wave functions of  $A=4$  $\Xi$ hypernuclei. }
\label{tab:Probcore_indiff_partialwave_a4}
\end{table*}

   \begin{table*}
    \renewcommand{\arraystretch}{2.0}
 \vskip 1 cm
\begin{center}
   \setlength{\tabcolsep}{0.085cm}

\begin{tabular}{|c|c |}
\cline{1-2}
$(J_{core},T_{core}, m^t_{\Xi})$ &   $^5_{\Xi}\mathrm{H}(\frac{1}{2}^+, \frac{1}{2})$ \\ 
\hhline{  = = }
$ (0, 0, -\frac{1}{2})$  & 96.03   \\ \hhline{  - - }
$ (0, 0, \frac{1}{2})$  & 1.1   \\ \hhline{  - - }
$ (0, 1, \frac{1}{2})$  & 2.1   \\ \hhline{  = = }
$ \text{others}$  & 0.3   \\  
\cline{1-2}
\end{tabular}
\hspace{1cm}
\begin{tabular}{|c|c |}
\cline{1-2} 
$(J_{core},T_{core}, m^t_{\Xi})$ & $^7_{\Xi}\mathrm{H}(\frac{1}{2}^+,\frac{3}{2})$ 
        \\ \hhline{  = = }
$ (0, 1, -\frac{1}{2})$  & 94.44 \\ \hhline{  - - }
$ (0, 2, -\frac{1}{2})$  & 0.7 \\ \hhline{  - - }
$ (0, 2, \frac{1}{2})$  & 2.8 \\  \hhline{  = =  }
$ \text{others}$  & 1.13  \\  
\cline{1-2} 
\end{tabular}
\end{center}
\caption{ Same as Table~\ref{tab:Probcore_indiff_partialwave_a4} for $A=5$ and 7  $\Xi$ hypernuclei. }
\label{tab:Probcore_indiff_partialwave_a57}
\end{table*}

\bibliographystyle{unsrturl}

\bibliography{./bib/hyp-literatur.bib,./bib/ncsm.bib,./bib/hypernuclei.bib,./bib/nn-interactions.bib,./bib/yn-interactions.bib,./bib/srg.bib,./bib/double-strangeness.bib, ./bib/Xi-system.bib}

\end{document}